\documentclass[aps,twocolumn,showpacs,preprintnumbers,amssymb]{revtex4}

\newcommand{\be}{\begin{eqnarray}}
\newcommand{\ee}{\end{eqnarray}}

\newcommand{\eins}{\mbox{$1 \hspace{-1.0mm}  {\bf l}$}}
\def\bea{\begin{eqnarray}}
\def\eea{\end{eqnarray}}
\def\C{\hbox{$\mit I$\kern-.7em$\mit C$}}
\def\N{\hbox{$\mit I$\kern-.3em$\mit N$}}

\def\tr{{\rm tr}}

\usepackage{dcolumn} 
\usepackage{bm} 

\usepackage{epsfig}

\usepackage{amsmath,amsfonts,amssymb,graphics,graphicx,epsfig,color,times,bbm}

\begin{document}



\title{Entanglement in spin chains and lattices with long--range Ising--type interactions}

\author{W. D{\"u}r$^{1}$, L. Hartmann$^{1}$, M. Hein$^{1}$, M. Lewenstein$^2$
and H.-J. Briegel$^{1,3}$}

\affiliation{$^1$ Institut f{\"u}r Theoretische Physik, Universit{\"a}t Innsbruck,
Technikerstra{\ss}e 25, A-6020 Innsbruck, Austria\\
$^2$ Institut f\"ur Theoretische Physik, Universit\"at Hannover, Appelstra{\ss}e 2, 
D-30167 Hannover, Germany\\
$^3$ Institut f\"ur Quantenoptik und Quanteninformation der \"Osterreichischen Akademie der Wissenschaften, Innsbruck, Austria.}
\date{\today}

\begin{abstract}
We consider $N$ initially disentangled spins, embedded in a ring or $d$-dimensional lattice of arbitrary geometry, which interact via some long--range Ising--type interaction. We investigate relations between entanglement properties of the resulting states and the distance dependence of the interaction in the limit $N \to \infty$. We provide a sufficient condition when bipartite entanglement between blocks of $L$ neighboring spins and the remaining system saturates, and determine $S_L$ analytically for special configurations. We find an unbounded increase of $S_L$ as well as diverging correlation and entanglement length under certain circumstances. 
For arbitrarily large $N$, we can efficiently calculate all quantities associated with reduced density operators of up to ten particles.

\end{abstract}


\maketitle


The investigation of entanglement properties of strongly interacting many body systems has proven to be a fruitful approach. Entanglement was shown to indicate quantum phase transitions \cite{Os02,Ost02,Vi03,Ve03} and the presence of long--range correlations even in systems with gapped Hamiltonians \cite{Ve03b}. In density matrix renormalization group (DMRG), a powerful numerical method which allows one to treat spin chains of up to a few hundred particles with high accuracy, the investigation of the role of entanglement has allowed one to understand \cite{Vi04,Sc05} and overcome \cite{Ve042d} limitations of the method. Standard DMRG can treat spin systems with a bounded amount of bipartite entanglement, measured by the entropy of entanglement $S_L$ between blocks of $L$ neighboring spins and the remaining systems (e.g. spin chains with short range interactions). The generalized method of Ref. \cite{Ve042d} can also handle spin systems arranged on two (and higher) dimensional lattices, where $S_L$ scales with the surface of the block, $S_L \propto L^{1/2}$.


Despite these significant developments, many spin systems in two and three dimensional setups remain untractable, among them disordered systems or systems with long range interactions where $S_L \propto L$. Such spin lattice systems with long--range interactions occur naturally in several quantum optical setups. There, Ising--type interactions are induced by other interactions with a characteristic distance dependence. Examples are the internal states of neutral atoms in an optical lattice that interact via an induced dipole interaction \cite{Br99} (see also \cite{Ku02}), or ions stored in microtraps where interactions are induced by pushing the ions dependent on their internal state such that they feel a different Coulomb potential \cite{Ci00,Po04,Ja02}. 
In this letter, we analyze spin chains and spin lattices in arbitrary dimensions with long--range Ising--type interactions. Despite the failure of known methods, the restriction to Ising--type interactions allows us to analyze both static and dynamical entanglement properties of the system in great detail. We find relations between entanglement properties of states $|\Psi_t\rangle$ ---resulting from evolution of the system initially prepared in some product state $|\Psi_0\rangle$ under the Hamiltonian $H$ for time $t$--- and the distance dependence of the interaction. We obtain information about the {\em dynamics} of entanglement and, at the same time, a large family of multipartite entangled states with rich entanglement properties. 

Our results are based on a description of the states $|\Psi_t\rangle$ in terms of generalized Valence Bond Solids (VBS) \cite{Ve04}. In this picture, we can efficiently calculate the reduced density operators of a small number $L\leq 10$ of arbitrary spins, even if the total number of spins $N$ is large (a standard PC can easily handle $N=10^5$ particles). We can hence determine all quantities associated with reduced density operators of small subsystems, including e.g. higher order correlation functions or (bounds on) bipartite entanglement $S_L$. We emphasize that for general pure states, the calculation of reduced density operators is a highly non-trivial task due to the exponential scaling with the system size $N$. 
For certain distance laws, we can describe the scaling of block--wise entanglement with the size of the block $L$ in the limit $N\to \infty$. In one dimension, we obtain a sufficient condition when block--wise entanglement saturates, which is the case whenever the distance dependence of the interaction strength scales as $1/r^\alpha$ with $\alpha>1$. For special configurations, we calculate $S_L$ analytically for all $L$ and show that $S_L$ can indeed grow unboundedly and proportional with $L$. This is in contrast to entanglement properties of 1D-VBS states recently analyzed in \cite{Fa04}, where $S_L$ is bounded by two. Finally we find that the correlation length diverges under certain circumstances, even if $S_L$ saturates.

In our model, we consider $N$ spin 1/2 systems (qubits) with pairwise interactions, described by an Ising--type Hamiltonian
\be\label{InteractionHamiltonian}
H=\sum_{k < l} f(k,l) \frac{1}{4}(\eins-\sigma_z^{(k)})\otimes(\eins-\sigma_z^{(l)}).
\ee
We assume that the spins are arranged on a $d$--dimensional lattice with fixed geometry and are initially completely polarized in $x$--direction, i.e. $|\Psi_0\rangle = |+\rangle^{\otimes N}$, where $|+\rangle =1/\sqrt{2}(|0\rangle + |1\rangle)$. The methods we develop can also describe disordered systems with random coefficients $f(k,l)$ and can take arbitrary (product) input states into account. We are interested in (entanglement) properties of the state 
\be\label{UnitaryEvolution}
|\Psi_t\rangle \equiv e^{-itH}|\Psi_0\rangle.
\ee
We consider the situation where the coupling between spins obeys a certain distance law, in the sense that the coefficients $f(k,l)$, describing the strength of the coupling, only depend on the distance $r_{kl}\equiv \|k-l\|$ between particles $k$ and $l$, $f(k,l)=f(r_{kl})$. In the example of ions stored in microtraps \cite{Ci00,Ja02} one finds for instance $f(r_{kl})=r_{kl}^{-3}$ \cite{Ja02}. 



{\bf Description in terms of Valence Bond Solids:} 
The unitary evolution operator $U(t) \equiv e^{-itH}$ in (\ref{UnitaryEvolution}) can equivalently be described by a product $U=\prod_{k,l}U_{kl}$ of {\em commuting} controlled phase gates $U_{kl} \equiv diag(1,1,1,e^{i\phi_{kl}})_{kl}$ acting on pairs of qubits, where $\phi_{kl}=f(r_{kl})t$.
If $U$ acts on a completely polarized state $|\Psi_0\rangle =|+\rangle^{\otimes N}$ and $\phi_{kl}\in\{0,\pi\}$, the resulting states are graph states \cite{Rau01,He03}. Hence, we will refer to states with arbitrary $\phi_{kl}$ (produced from $|+\rangle^{\otimes N}$) as {\em weighted graph states}. Verstraete and Cirac recently proposed a description of graph states in terms VBS \cite{Ve04}. Instead of maximally entangled pairs of qubits, we use pairs where the degree of entanglement depends on $\phi_{kl}$. In this generalized VBS--picture, we develop a description of all weighted graph states and extend it to all states produced by the interaction Hamiltonian $H$ acting on any initially unentangled state.

Each qubit $k$ of a physical state $|\Psi_t\rangle$ is replaced by $N-1$ (virtual) qubits $k_1\ldots k_{N-1}$. The VBS state $|\tilde \Psi_t\rangle$ with corresponding Hilbert space ${\cal H}=[(\C^2)^{N-1}]^N$ is given by a tensor product of $N(N-1)/2$ independent, non--maximally entangled pairs of qubits $|\chi_{k_il_j}\rangle = U_{k,l} |+\rangle_{k_i}|+\rangle_{l_j}$ shared between virtual qubits $k_i,l_j$ of parties $k,l$, where each $k_i,l_j$ appears only once. Up to a normalization factor, we obtain the corresponding weighted graph state $|\Psi_t\rangle=U|+\rangle^{\otimes N}$ from $|\tilde \Psi_t\rangle = \bigotimes |\chi_{k_il_j}\rangle$ by performing local projections $P_k=|0_k\rangle\langle \vec 0_{\vec k}|+|1_k\rangle\langle \vec 1_{\vec k}|$ onto two dimensional subspaces at all locations $k$, where $\vec k=k_1\ldots k_{N-1}$ and $|\vec m\rangle =|mm\ldots m\rangle$. The VBS--like state $|\tilde \Psi_t\rangle$, together with the projection $\bigotimes_k P_k$, thus provides an equivalent description of the state $|\Psi_t\rangle$. We can generalize this description to arbitrary product input states $|\varphi_1\ldots \varphi_N\rangle$. In this case, the (unnormalized) VBS--like state is of the form $|\tilde \Psi_t\rangle=\bigotimes_{k,l}|\chi_{k_il_j}\rangle$ with $|\chi_{k_il_j}\rangle = U_{kl}|\sqrt[N-1]{\varphi_k}\rangle_{k_i}|\sqrt[N-1]{\varphi_l}\rangle_{l_j}$, where $|\varphi\rangle =\alpha|0\rangle+\beta|1\rangle$ and$|\sqrt[n]{\varphi}\rangle\equiv\sqrt[n]{\alpha}|0\rangle + \sqrt[n]\beta|1\rangle$. 
In the following, we use this description to determine reduced density operators $\rho_A$. For the sake of simplicity, we consider weighted graph states, i.e. states arising from input states $|+\rangle^{\otimes N}$.

We denote by $A$ an arbitrary subset of the $N$ qubits, and we call the set of remaining qubits $B$. Because all $U_{kl}$ commute and because unitaries in $B$ do not influence $\rho_A= \tr_B(|\Psi_t\rangle\langle\Psi_t|)$, we can write 
\be
\rho_A = \prod_{k,l\in A} U_{kl} \tr_B |\Psi_t'\rangle\langle\Psi_t'| U_{kl}^{\dagger},
\ee 
with $|\Psi_t'\rangle = \prod_{k\in A, l\in B} U_{kl} |\Psi_0\rangle$. We now determine $\rho_A'=\tr_B |\Psi_t'\rangle\langle\Psi_t'|$ in the VBS--picture, i.e. we start with the state 
\be
|\tilde \Psi'_t\rangle = \bigotimes_{k\in A,l\in B} U_{kl}|+\rangle_{k_i}|+\rangle_{l_j}.
\ee 
For the following argumentation it is crucial that $U_{kl} = \eins$ for $k,l \in B$ as is the case for $|\tilde\Psi'_t\rangle$ but not for $|\tilde \Psi_t\rangle$. Performing the projections $P_l$ on all particles $l \in B$ (but not in $A$) leaves us with a state of the form $\bigotimes_{l\in B} [\prod_{k\in A} U_{kl}|\vec +\rangle_{\vec k}|+\rangle_{l}]$. We have a tensor product of $|B|$ states where in the $l^{th}$ state particle $l \in B$ is entangled with a virtual system $A$. For each of these states we can independently calculate the reduced density operator with respect to $A$. We simply trace out particle $l$ and obtain 
\be
\rho'_A(l)=\textstyle\frac{1}{2}(|\vec +\rangle_A\langle \vec +| + |\chi_l\rangle_A\langle\chi_l|),
\ee 
with $|\chi_l\rangle_A=\bigotimes_{k\in A} (|0\rangle +e^{-i\phi_{kl}}|1\rangle)/\sqrt{2}$. Now, we perform the projections $P_k$ for all $k \in A$. The resulting density operator $\rho'_A$ is (up to normalization) given by the {\em Hadamard product} of all density operators $\rho'_A(l)$, where the Hadamard product of two matrices corresponds to component--wise multiplication in the computational basis. The reduced density operator $\rho_A$ is then obtained from $\rho_A'$ by taking into account interactions within $A$ leading to $\rho_A = \prod_{k,l\in A} U_{kl} \rho_A' U_{kl}^{\dagger}$. Finally, we must normalize the resulting state.

{\bf Computable quantities:} 
The method outlined above provides an efficient way to calculate reduced density operators for all weighted graph states. The computation time is linear in the number $|B|$ of particles in the remaining system (but exponential in $|A|$), as opposed to an exponential scaling in $N=|A|+|B|$ of computation time and memory cost for general pure states. Hence, for arbitrary large systems, all quantities that depend on the reduced density operator of a small number of qubits can be calculated efficiently. For instance, from $\rho_A$ of one and two qubits, we can determine all two-point (and also higher order) correlation functions $Q_{\alpha,\beta}^{k,l}$, 
\be
Q_{\alpha,\beta}^{k,l}=\langle\sigma_\alpha^{(k)}\sigma_\beta^{(l)}\rangle- \langle\sigma_\alpha^{(k)}\rangle\langle\sigma_\beta^{(l)}\rangle,
\ee 
lower and upper bounds on the localizable entanglement $E_L$ \cite{Ve03}, the entanglement of formation between pairs of particles, as well as the multipartite entanglement measure $E_{\rm MW}$ \cite{Me02}. The maximal classical correlation $Q_{\max}^{k,l}$ between two particles is given by the largest singular value of the matrix $Q_{\alpha,\beta}^{k,l}$ \cite{Ve03}. The localizable entanglement $E_L^{k,l}$ is the maximum amount of entanglement that can be established between a pair of particles $k,l$, on average, by performing local measurements on all other particles. The relation $Q_{\max}^{k,l} \leq E_L^{k,l} \leq E_A^{k,l}$ holds \cite{Ve03}, where $E_A^{k,l}$ is the concurrence of assistance \cite{La03}. The measure $E_{\rm MW}$ is given by $E_{\rm MW} = 2[1-1/N\sum_k \tr(\rho_k^2)]$ \cite{Me02}. In addition, the (bipartite) entanglement between blocks of a small number $L\leq 10$ of neighboring spins is measured by the entropy of entanglement $S_L$, that is the von Neumann entropy of the reduced density operator $\rho_L$, $S_L=\tr(\rho_L \log_2 \rho_L)$. Clearly, $0 \leq S_L \leq L$, where $S_L=L$ indicates maximal entanglement between the blocks. For blocks larger than 10 qubits, we make use of the strong subadditivity of the entropy to derive upper bounds on $S_L$. By breaking a block of size $L$ into $n$ sub-blocks $L_i$ of size $|L_i|=L/n$, we obtain 
\be
{\cal S}(\rho_L)\leq\sum_{i=1}^{n-1}{\cal S}(\rho_{L_i,L_{i+1}})-\sum_{i=2}^{n-1}{\cal S}(\rho_{L_i}),
\ee 
where $\rho_{L_i,L_{i+1}}$ is a sub-block of length $2L/n$. The unitary operations $U_{kl}, {k,l\in A}$, do not change the entropy $S(\rho_A)$ and hence the reduced density operator $\rho_A'$ can be used directly since $S(\rho_A)=S(\rho_A')$. Nevertheless, upper bounds on the entropy are different for $\rho_A$ and $\rho_A'$, where the latter turn out to be more stringent and will hence be used in the following.

The fact that the total reduced density operator is given by the Hadamard product of reduced density operators with respect to all particles in the system $B$ can be exploited to prove monotonicity properties of $S(\rho_A')$. For a fixed size $|A|$, we add one particle $j$ to $B$. The reduced density operator is updated by Hadamard multiplication with $\rho_A'(j)$. From theorem 5.5.12 in \cite{Ho91} follows that the eigenvalues of the resulting density operator are majorized by the eigenvalues of the initial one, which implies that the entropy increases \cite{Vi00}. As a consequence we obtain lower bounds on the entropy of entanglement $S(\rho_A')$ when we take into account only a subset $\tilde B \subset B$ of all particles (and ignore the other particles in $B$).

\begin{figure}[ht]
\begin{picture}(230,90)
\put(-5,0){\epsfxsize=230pt\epsffile[26 13 809 300]{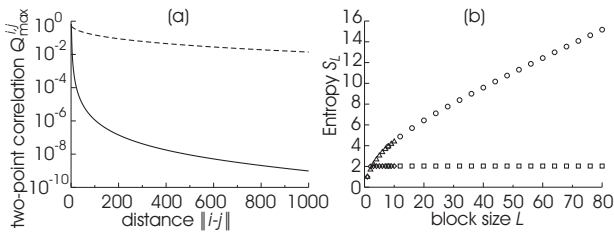}}
\end{picture}
\caption[]{\label{UpperBounds} Spin chain with $N=10^5$, $t=0.3 \pi$ and $f(r_{kl})=r^{-\alpha}$ for different $\alpha$. (a) Maximal two point correlation $Q_{\max}^{i,j}$ as a function of the distance $\|i-j\|$ for $\alpha=1/3$ (dashed) and $\alpha=3$ (solid). (b) Exact values [upper bounds] of entropy of entanglement $S_L$ as function of block size $L$ for $\alpha=1/3$ ($\vartriangle[\circ]$) and $\alpha=3$ ($\diamond[\square]$).}
\end{figure}

{\bf Static properties of resulting states:}
We apply these results to determine (static) entanglement properties of the state $|\Psi_t\rangle$ for some fixed time $t<\pi$ and for different distance laws $f(r_{kl})=r_{kl}^{-\alpha}, \alpha>0$. 
Figure \ref{UpperBounds}(a) shows the maximal two--point correlation $Q_{\max}^{i,j}$ in a chain of $N=10^5$ particles as a function of of the distance between particles $\|i -j\|$. We observe that correlations decay slower than exponential. Therefore, the correlation length $\xi$ and also the entanglement length $\xi_E$ {\em diverge} \cite{Ve03}. This indicates long--range quantum correlations for all power laws, as we find that only exponential fall--off functions $f(k,l)=e^{-\kappa r_{kl}}$ lead to a finite correlation length. 
Figure~\ref{UpperBounds}(b) shows the scaling of the entropy of entanglement $S_L$ with the block size $L$ for different power laws. Exact values are plotted for $L\leq 10$, while upper bounds (corresponding to $|L_i|=4$) are plotted for $L\ge 10$. The upper bound on $S_L$ seems to grow unboundedly for $\alpha \leq 1/2$, whereas $S_L$ saturates for $\alpha > 1$. For $\alpha>1$ the system thus contains a bounded amount of entanglement $S_L$, but has a diverging correlation length $\xi$.

The saturation of $S_L$ for $\alpha > 1$ can be proven analytically when we take the limit of an infinite chain ($N\to \infty$) and afterwards let the block size also go to infinity. Both steps involve infinite products and sums, and we must concern ourselves with convergence/divergence of these products and sums. In an infinite chain, the reduced density operator for one qubit is $\rho_A =\frac{1}{2} [\eins + c|0\rangle\langle 1| + c^*|0\rangle\langle 1|]$ with $c=\prod_{k=1}^\infty\cos\frac{\phi_{1k}}{2}e^{-i\sum_{k=1}^\infty\phi_{1k}/2}$ and $\phi_{1k}=r_{1k}^{-\alpha} t$ \cite{Va04}. The eigenvalues of $\rho_A$ are given by $(1\pm |c|)/2$, so we can omit the phase of $c$. We write $\prod_{k=1}^\infty\cos \phi_{1k}=e^{\sum_{k=1}^\infty\ln\cos \phi_{1k}}$. Taylor series expansions and Cauchy's integral criterion tell us that the sum in the exponent converges for $\alpha>1/2$ and diverges to $-\infty$ for $0<\alpha\leq1/2$. Hence, the infinite product is non-zero in the first case and zero in the second case. The entropy of entanglement $S_A$ is thus smaller than one for $\alpha>1/2$ and equals one for smaller $\alpha$. In other words, for slow fall--off functions (strong long--range interactions), the entropy of a single particle is maximal, independent of the time $t>0$, as infinitely remote regions still influence the qubit we consider.


To take the limit $L\to\infty$ in the case $\alpha>1/2$ we use a bound given by $S(\rho_L)=S(\rho_L') \leq \sum_{i=1}^N S(\rho_{L_i}')$ for sub-block sizes $|L_i|=1$. In this case, the convergence properties of the infinite sum of single particle entropies (each itself given by an infinite product) can be determined by using again Taylor series expansion and Cauchy's integral criterion. In the limit $N\to \infty$ and $L \to \infty$, the upper bound converges to a constant value for $\alpha >1$ and hence the exact value of the entropy $S_L$ also saturates as a function of $L$ for power laws $f(r_{kl}) = r_{kl}^{-\alpha}$ with $\alpha>1$. We can generalize this result to $d$--dimensional lattices. When considering blocks of $L$ particles contained in a $d$--dimensional ball, $S_L$ can at most grow like the volume of that ball, whereas we find that for $\alpha > d$ the upper bound on $S_L$ grows at most like the surface of the ball. The proof is similar to the one--dimensional case.

For special cases, we get a complete analytic description of the entanglement properties. In the following, we consider an interaction Hamiltonian with a fixed interaction length $\lambda$ and constant interaction strength, i.e. $f(r_{kl}) = 1$ if $r_{kl} \leq \lambda$ and zero otherwise. For $t=\pi$, the resulting states $|\Psi_{\pi}\rangle$ are special instances of graph states \cite{Rau01}. We denote a $d$-dimensional quadratic block of size $L=a^d$ neighboring spins by $A$ and the remaining system by $B$. We measure the bipartite entanglement between $A$ and $B$ by the entropy of entanglement $S_L$. For graph states, $S_L$ is given by the binary rank of the adjacency matrix $\Gamma_{AB}$ between the quadratic block $A$ and the rest $B$ \cite{He03}.
If the lattice is large enough to contain not only the block $A$ but also the larger block of size $(a+2\lambda)^d$ with $A$ in its center, then no boundary effects have to be taken into account. We can inductively show that the matrix $\Gamma_{AB}$ has maximal rank by considering different layers $A_k$ with geometric distance $k$ to $B$, starting with $k=1$. Hence, $S_L$ is simply given by the number of vertices within $A$ that are connected to the rest $B$, so $S_L=a^d-[a-\min(2\lambda,a)]^d$ with $a=\sqrt[N]{L}$. 
In a general situation, no such simple rule to calculate $S_L$ holds (in contrast to what is suggested in \cite{Ve04}). A counterexample is given by a state with $N=4$, $\varphi_{13}=\varphi_{14}=\varphi_{23}=\varphi_{24}=\pi$, which has $S(\rho_{12})=1$. For an (infinite) chain of particles we obtain $S_L=\min(2\lambda,L)$. As $L$ increases, $S_L$ saturates at the value $2\lambda$ for any fixed interaction length $\lambda$. Only if $\lambda$ itself goes to infinity as $N \to \infty$, $S_L$ can grow (unboundedly) with $L$ when the ratio of the interaction length to the total number of particles is kept constant. For a given interaction length $\lambda$, and for any ball containing $L$ particles, only those particles that have connections to the remaining system contribute to $S_L$. The entropy $S_L$ equals $L$ if the radius of the hypersphere is smaller than $\lambda$. Otherwise, $S_L$ scales essentially like the volume of a surface shell with thickness $\lambda$ , that is $S_L \propto \lambda L^{(d-1)/d}$. Two--point correlation functions $Q_{\alpha,\beta}^{i,j}$ turn out to be zero for all pairs of particles (except for the case of two neighboring particles at the end of a chain), because their reduced density matrix $\rho_{kl}$ is the identity \cite{He03}. Nevertheless, $|\Psi_\pi\rangle$ is maximally connected \cite{Rau01}, which means that a Bell state between any pair of particles can be obtained by local measurements on the rest of the particles. Hence $|\Psi_\pi\rangle$ has maximal localizable entanglement between any two particles, $E_L=1$, and, thus, an infinite entanglement length. Finally, $E_{MW}=1$ for all such states since the reduced density operator of each single particle is also the identity.

{\bf Dynamics of entanglement:}
We return to the case of a chain of particles and to Hamiltonians with arbitrary distance dependence $f(k,l)=r_{kl}^{-\alpha}$ and consider the {\em dynamics} of entanglement, that means the change of entanglement and correlations of the state $|\Psi_t\rangle$ with time. The scaling of the entropy with block size $L$ is essentially still governed by the the specific form of the distance dependence for any finite $t$, because infinitely remote regions still influence a subsystem in a similar way as discussed before.

\begin{figure}[ht]
\begin{picture}(230,90)
\put(-5,0){\epsfxsize=230pt\epsffile[28 22 844 322]{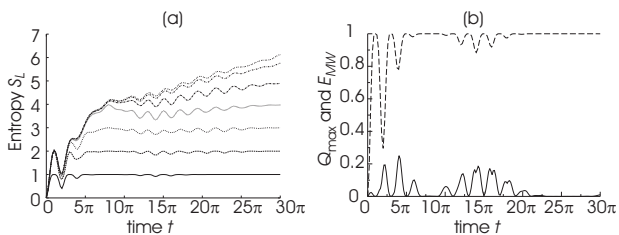}}
\end{picture} 
\caption[]{\label{CorrAssistanceEntropy}Dynamics of entanglement for chain with $N=10^5$ and $f(r_{kl})=r_{kl}^{-3}$. (a) Entropy of entanglement $S_L$ for blocks $L=1$ (bottom) up to $L=7$ (top). 
(b) Two-point correlations $Q_{\max}^{i,j}$ for $\|i-j\|=5$ (solid) and $E_{\rm MW}$ (dashed).}
\end{figure}

For large times $t$, more and more of the interaction phases $\phi_{kl}=f(k,l)t$ start to oscillate (as they are effectively taken modulo $\pi$) and approach in the limit of large $t$ a (quasi)--random distribution. In the limit of an infinite chain and $t \to \infty$, the entropy of the reduced density operator of any finite group $A$ of particles is maximal, $S(\rho_A)=|A|$. This can be seen by considering the off diagonal elements of reduced density operators, which all contain infinite products of cosines of (sums of) {\em random} angles. All these products tend to zero for $N\to \infty$, leading to a maximally mixed state.
For a chain of $N=10^5$ particles with $f(r_{kl})=r_{kl}^{-3}$ the time dependence of the entropy of entanglement for blocks up to size $7$ is plotted in Fig.~\ref{CorrAssistanceEntropy}(a), while Fig.~\ref{CorrAssistanceEntropy}(b) shows two point correlation functions $Q_{\max}$ and the multipartite measure $E_{MW}$.

In this paper, we have investigated entanglement properties of states generated from product states by long--range Ising--type interactions. For an (arbitrary) total number of particles $N$, and using a description in terms of generalized VBS, we could efficiently determine the reduced density operators of a few ($\le 10$) particles and hence all quantities which are associated with reduced density operators. 
For different distance laws, we investigated the scaling of block--wise entanglement and showed that in 1D $S_L$ saturates for $f(r_{kl}) \propto r_{kl}^{-\alpha}$ for all $\alpha >1$. We also found diverging correlation- and entanglement lengths for all power laws. Our methods can also be applied to disordered systems, such as quantum lattice gases, spin glasses, or the semi quantal Boltzmann gas (introduced in \cite{He03}); they can be extended to describe the dynamics of arbitrary slightly entangled input states (in the sense of small Schmidt measure \cite{Br00}) under this Ising--type interaction. We remark that related studies have recently been performed for harmonic lattice systems in Ref. \cite{Pl04}.

We thank J.I. Cirac, S.J. van Enk and F. Verstraete for discussions. This work was supported by the \"OAW 
through project APART (W.D.), the European Union (IST-2001-38877,-39227) and the Deutsche Forschungsgemeinschaft.


\end{document}